\documentclass[sn-mathphys,Numbered]{sn-jnl}% Math and Physical Sciences Reference Style
\usepackage{graphicx}%
\usepackage{multirow}%
\usepackage{amsmath,amssymb,amsfonts}%
\usepackage{amsthm}%
\usepackage{mathrsfs}%
\usepackage[title]{appendix}%
\usepackage{xcolor}%
\usepackage{textcomp}%
\usepackage{manyfoot}%
\usepackage{booktabs}%
\usepackage{algorithm}%
\usepackage{algorithmicx}%
\usepackage{algpseudocode}%
\usepackage{listings}%
\usepackage{caption}
\usepackage{hyperref}

\usepackage{tabularx}
\usepackage{adjustbox}
\usepackage{setspace} % 加载 setspace 宏包
\usepackage{geometry}
\hypersetup{
	colorlinks=true,
	linkcolor=black,
}
\usepackage{siunitx}
\usepackage{lineno}
\usepackage{setspace}
\makeatletter

\makeatletter
\usepackage[skip=15pt]{caption}
\usepackage{siunitx}%
\usepackage{indentfirst}% 
\usepackage[version=4]{mhchem}%
\usepackage{natbib}
\setcitestyle{super,open={},close={}}
\makeatletter
\renewcommand{\@biblabel}[1]{#1.}

\makeatother
\theoremstyle{thmstyleone}%
\theoremstyle{thmstyletwo}%
\theoremstyle{thmstylethree}%
\begin{document}
	\renewcommand{\thelinenumber}{\normalfont\fontsize{10}{12}\selectfont\arabic{linenumber}} %Renew the line number command to change font size and style
%\linenumbers
\title[Article Title]{Synthetic frequency-controlled gene circuits unlock expanded cellular states  }

\author[1,2]{\vspace*{5pt}\fnm{Rongrong Zhang} \sur{}}
\equalcont{Rongrong Zhang and Shengjie Wan contributed equally to this work.}
\author[1,2]{\fnm{Shengjie Wan} \sur{}}
\equalcont{Rongrong Zhang and Shengjie Wan contributed equally to this work.}

\newpage
\author[3,1]{\fnm{Jiarui Xiong} \sur{}}
\author[1,2]{\fnm{Lei Ni} \sur{}}
\author[1,2]{\fnm{Ye Li} \sur{}}
\author[1,2]{\fnm{Yajia Huang} \sur{}}
\author[1,2]{\fnm{Bing Li} \sur{}}
\author[1,2]{\fnm{Mei Li} \sur{}}
\author*[4,5]{\fnm{Shuai Yang} \sur{}}
\author*[1,2]{\fnm{Fan Jin} \sur{}}\email{Fan Jin, fan.jin@siat.ac.cn; Shuai Yang, yangs@clas.ac.cn}

\affil[1]{\orgdiv{CAS Key Laboratory of Quantitative Engineering Biology, Shenzhen Institute of Synthetic Biology, Shenzhen Institute of Advanced Technology}, \orgname{Chinese Academy of Sciences}, \orgaddress{\street{} \city{Shenzhen}, \postcode{518055}, \state{} \country{China}}\vspace*{5pt}}
\affil[2]{\orgdiv{Shenzhen Synthetic Biology Infrastructure, Shenzhen Institute of Synthetic Biology, Shenzhen Institutes of Advanced Technology}, \orgname{Chinese Academy of Sciences}, \orgaddress{\street{} \city{Shenzhen}, \postcode{518055}, \state{} \country{China}}\vspace*{5pt}}
\affil[3]{\orgdiv{Hefei National Research Center for Physical Sciences at the Microscale, Department of Polymer Science and Engineering}, \orgname{University of Science and Technology of China}, \orgaddress{\street{} \city{Heifei}, \postcode{230026}, \state{} \country{China}}\vspace*{5pt}}
\affil[4]{\orgdiv{National Science Library (Chengdu)}, \orgname{Chinese Academy of Sciences}, \orgaddress{\street{} \city{Chengdu}, \postcode{610299}, \state{} \country{China}}\vspace*{5pt}}
\affil[5]{\orgdiv{Department of Information Resources Management, School of Economics and Management}, \orgname{ University of Chinese Academy of Sciences}, \orgaddress{\street{} \city{Beijing}, \postcode{100049}, \state{} \country{China}}\vspace*{5pt}}

\maketitle

% 强制换页，将摘要放到新页
\newpage

% 调整摘要部分页边距
\newgeometry{left=35mm, right=35mm, top=30mm, bottom=30mm} % 适度增大页边距
% 摘要部分
\begin{center}
	\textbf{\large Abstract} % 居中加粗显示摘要标题
\end{center}

\vspace{10pt} % 添加空白以区分标题和正文

\noindent Natural biological systems process environmental information through both amplitude and frequency-modulated signals, yet engineered biological circuits have largely relied on amplitude-based regulation alone. Despite the prevalence of frequency-encoded signals in natural systems, fundamental challenges in designing and implementing frequency-responsive gene circuits have limited their development in synthetic biology. Here we present a Time-Resolved Gene Circuit (TRGC) architecture that enables frequency-to-amplitude signal conversion in engineered biological systems. Through systematic analysis, we establish a theoretical framework that guides the design of synthetic circuits capable of distinct frequency-dependent responses, implementing both high-pass and low-pass filtering behaviors. To enable rigorous characterization of these dynamic circuits, we developed a high-throughput automated platform that ensures stable and reproducible measurements of frequency-dependent r esponses across diverse conditions. Using this platform, we demonstrate that these frequency-modulated circuits can access cellular states unreachable through conventional amplitude modulation, significantly expanding the controllable gene expression space in multi-gene systems. Our results show that frequency modulation expands the range of achievable expression patterns when controlling multiple genes through a single input, demonstrating a new paradigm for engineering cellular behaviors. This work establishes frequency modulation as a powerful strategy for expanding the capabilities of engineered biological systems and enhancing cellular response to dynamic signals.

\vspace{10pt} % 添加空白以区分摘要正文和关键词

\restoregeometry

\section*{Main}
Information theory fundamentally shapes our understanding of how complex systems process and transmit signals. Central to this framework are two distinct approaches to information encoding: Amplitude Modulation (AM) and Frequency Modulation (FM)\cite{roder1931amplitude}. While these complementary strategies have been extensively developed in engineered systems, their implementation in biological contexts reveals an intriguing asymmetry. Natural biological systems routinely employ both modalities\cite{hao2012signal,purvis2013encoding,li2019communication,cai2008frequency}, yet synthetic biology has predominantly focused on amplitude-based regulation, leaving the potential of frequency-based control largely unexplored\cite{nandagopal2011synthetic,levine2013functional,luo2024quantitative}.

The sophistication of natural genetic regulatory systems reflects billions of years of evolution, yielding intricate networks that process environmental information through both amplitude and frequency-encoded signals\cite{micali2015accurate,szischik2024transient}. These networks demonstrate remarkable precision in maintaining cellular homeostasis and coordinating complex responses to environmental cues. Frequency-based regulation appears particularly crucial in various natural contexts, from calcium signaling\cite{cai2008frequency} to hormone secretion patterns\cite{brabant1992pulsatile} and transcription factor dynamics\cite{li2018frequency,behar2010understanding,locke2011stochastic,hao2012signal,crabtree2008bursting}. The prevalence of frequency modulation in natural systems suggests its fundamental importance in biological information processing\cite{purvis2013encoding}.

Synthetic biology has provided fundamental insights into gene circuit design principles, advancing from single-gene manipulation to the engineering of complex genetic circuits\cite{english2021designing,lim2010designing,mukherji2009synthetic}. The field has developed diverse tools for regulating cellular protein levels, primarily through amplitude modulation (AM) where gene expression is controlled by adjusting input signal strength. This approach, exemplified by hormone and light-inducible transcription factor systems\cite{romano2021engineering,baumschlager2021synthetic}, typically results in graded and sustained activation of transcriptional regulators. 
Building on amplitude-modulated regulation and integrating diverse physical and mathematical approaches to model biochemical dynamics, the field has achieved significant milestones including the construction of \cite{gardner2000toggleswitch}, oscillators\cite{stricker2008oscillator,elowitz2000synthetic}, and logic gates\cite{jones2022genetic,moon2012genetic,karlebach2008modelling}.

However, recent studies of natural systems reveal that many regulatory proteins, particularly transcription factors, exhibit sophisticated temporal activation patterns that go beyond simple amplitude modulation\cite{levine2013functional}. While synthetic biology has begun to explore dynamic signal regulation through optogenetic systems, these efforts have largely focused on pulse-width modulation (PWM)\cite{davidson2013programming,benzinger2018pulsatile,benzinger2022synthetic,ravindran2022synthetic}, where quantitative information is still encoded through time-averaged signal intensity. This represents an important step toward dynamic control but does not fully capture the rich frequency-modulated dynamics observed in natural systems\cite{bettenworth2022frequency,hao2012signal,albeck2013frequency,paszek2010oscillatory,berridge1997and}, where information is encoded purely in the frequency of transitions between "on" and "off" states while maintaining constant time-averaged signals. This limited exploitation of true frequency modulation in synthetic circuits represents a significant gap between engineered and natural biological systems, potentially constraining our ability to design more sophisticated cellular behaviors.

The development of frequency-modulated synthetic circuits has faced two major challenges. First, the absence of a comprehensive theoretical framework has limited our ability to understand and rationally design frequency-based regulation. Second, practical implementation requires precise temporal control of inputs and accurate measurement of dynamic responses - capabilities that have only recently become technically feasible\cite{zhang2023high,delvigne2023advances,foncillas2024automated}.
Here we present the Time-Resolved Gene Circuit (TRGC) architecture, a systematic framework for implementing frequency-based regulation in synthetic gene circuits. At its core is the Frequency-Amplitude Converter (FAC), a design that transforms frequency-encoded signals into defined gene expression patterns. By converting frequency-modulated inputs into amplitude-modulated outputs, the FAC creates bridges the gap  between frequency-based regulation and existing amplitude-dependent gene circuits, enabling the integration of dynamic signal processing with established synthetic biology tools.

Our work establishes comprehensive foundations for frequency-modulated gene circuits through three complementary advances. First, we developed an analytical framework that reveals the fundamental principles of cellular frequency-to-amplitude signal conversion, providing quantitative insights into how cells can process frequency-encoded information. Second, guided by these theoretical principles, we engineered synthetic circuits using optogenetic controls that implement precise frequency-dependent regulation. These circuits demonstrate remarkable versatility, capable of operating as either high-pass or low-pass frequency filters depending on their configuration, enabling selective responses to different frequency ranges. Third, to rigorously characterize these dynamic circuits, we developed a high-throughput automated platform that enables systematic exploration of frequency-dependent behaviors across diverse conditions. This integrated approach of theory, design, and validation creates a robust foundation for engineering frequency-responsive gene circuits.

This work reveals that frequency modulation significantly expands the achievable range of gene expression states in multi-gene systems. Our results demonstrate that frequency-based control can access cellular states inaccessible through conventional amplitude modulation alone, suggesting new possibilities for engineering complex cellular behaviors. These findings not only advance our capability to design sophisticated synthetic circuits but also provide insights into how natural systems may exploit frequency-based regulation for information processing.

By establishing a theoretical and experimental framework for frequency-modulated gene circuits, this study opens new avenues for engineering cellular behavior and understanding temporal information processing in biological systems. The principles developed here could enable novel applications in biotechnology and medicine while deepening our understanding of natural regulatory systems.

\subsection*{Frequency Amplitude Converter (FAC) Architecture Framework and Gene Circuit Design }
To achieve frequency-based regulation of gene expression, we needed to develop a system capable of converting frequency-encoded signals into defined protein expression levels. This conversion presents three fundamental challenges: First, the system must respond rapidly to dynamic input signals while maintaining temporal precision. Second, it must selectively process signals based on their frequency characteristics to enable frequency discrimination. Finally, it must integrate these processed signals over time to produce stable protein expression levels.

Analysis of these requirements led us to design a three-module architecture:  a Wave Converter (M1) for rapid signal processing, a Thresholding Filter (M2) for frequency-selective response, and an Integrator (M3) for temporal averaging. We termed this architecture the Frequency Amplitude Converter (FAC) (Fig. 1a). The Wave Converter transforms input signals into sawtooth wave patterns that fluctuate between peak ($s_\text{H}$) and trough ($s_\text{L}$) values, providing the dynamic response necessary for frequency detection. The Thresholding Filter then selectively processes these fluctuating signals based on defined thresholds, enabling frequency discrimination. Finally, the Integrator performs temporal averaging of the filtered signals to generate stable protein expression outputs. Through this sequential processing, the FAC effectively converts frequency-encoded inputs into defined amplitude outputs (Fig. 1b).

The frequency discrimination capabilities of this architecture emerge from the distinct behaviors of each module. The Wave Converter generates characteristic sawtooth waveforms that vary with input frequency while maintaining constant duty cycle and amplitude. Notably, low-frequency inputs produce waveforms with longer periods and larger amplitude fluctuations, characterized by higher peak values and lower trough values  compared to high-frequency inputs.

The system's frequency selectivity is primarily determined by the threshold setting ($s^*$) of the Thresholding Filter. With high threshold settings (Fig. 1c, upper panel), the filter preferentially blocks high-frequency signals while partially transmitting low-frequency signals, resulting in a Low-Pass FAC configuration. Conversely, low threshold settings (Fig. 1c, lower panel) enable complete transmission of high-frequency signals while partially filtering low-frequency signals, creating a High-Pass FAC configuration.
To implement this FAC architecture in biological systems, we developed the Time-Resolved Gene Circuit (TRGC) with strategically decoupled modules\cite{del2016control}. The key to successful implementation lay in establishing appropriate temporal separation between modules, enabling sequential signal processing while minimizing interference. This temporal hierarchy relies on distinct characteristic time constants ($T_{c1}$, $T_{c2}$, $T_{c3}$) for each module.

Our design process began with systematic analysis of these time constants, leading to the development of a Chemical Reaction Network (CRN) model (Supplementary Fig. 1). This framework enabled both temporal signal simulation and systematic exploration of dynamic interactions within the circuit (Supplementary Tables 1-3). The Integrator functions as the terminal output component, where protein expression integrates and averages signals from activated promoters. This process, limited by protein degradation rates, operates with characteristic time $T_{c3}$ ranging from $10^3$ to $10^4 $ seconds\cite{kaern2005stochasticity}.

The Thresholding Filter implements signal processing through two Hill kinetic processes: formation of transcription factor-signal molecule complexes and subsequent promoter activation. These rapid chemical reactions operate with characteristic time  $T_{c2}$ ($10^{-2}$ to $10^0$ seconds)\cite{mazzocca2021transcription}, significantly faster than  $T_{c3}$.

The Wave Converter bridges these timescales, with its characteristic time $T_{c1}$ positioned between $T_{c2}$ and $T_{c3}$. We implemented M1 using optogenetic components, specifically light-sensitive adenylyl cyclase (bPAC) and phosphodiesterase (CpdA), enabling precise control of cAMP levels through light stimulation. The system's response dynamics, primarily determined by CpdA-mediated cAMP degradation, exhibit characteristic time  $T_{c1}$ between $10^0$ and $10^1$ seconds\cite{stierl2011light}, satisfying $T_{c2} \ll T_{c1} \ll T_{c3}$.

This temporal hierarchy enables efficient signal processing: the Wave Converter translates light signals into cAMP dynamics, the Thresholding Filter rapidly processes these signals through Vfr-mediated transcriptional regulation, and the Integrator captures the cumulative response through protein expression. The resulting TRGC architecture provides a robust platform for implementing frequency-based gene regulation in biological systems.

%Our study aimed to investigate the information transmission capacity of the cAMP signaling pathway in \textit{P. aeruginosa}, a bacterium renowned for its adaptability and involvement in various cellular processes\cite{strateva2009pseudomonas}. To isolate and quantify the information transmission capabilities of cAMP as a signaling molecule, we needed to create a simplified, controllable system that would allow us to study the cAMP channel in isolation from its complex regulatory network. To achieve this, we first identified key proteins in the cAMP signaling pathway of \textit{P. aeruginosa} (Fig. 1a). Adenylyl cyclases CyaA and CyaB\cite{topal2012crystal}, catalyze the synthesis of cAMP from ATP, while the Vfr protein acts as a cAMP-dependent transcription factor\cite{fuchs2010pseudomonas}, transmitting the signal to downstream targets. These components form a bow-tie structure in the signaling pathway, with cAMP at the center.

\subsection*{Theoretical Modeling and Analysis of TRGC}
To analyze the TRGC system, we developed two complementary theoretical frameworks. A Chemical Reaction Network (CRN) model provided detailed simulation capabilities but its complexity made parameter-output relationships difficult to extract\cite{van2015programmable}. We therefore developed an analytical framework focusing on essential dynamics, enabling direct analysis of key parameters while maintaining biological realism\cite{brophy2014principles,kaern2005stochasticity}. This simplified approach bridges detailed simulations and experimental implementation.

The analytical framework exploits natural temporal hierarchies, where the Thresholding Filter (M2), Wave Converter (M1), and Integrator (M3) operate on millisecond-to-second, second-to-minute, and minute-to-hour timescales, respectively. This temporal separation ($T_{c2} \ll T_{c1} \ll T_{c3}$) enables modular analysis, leading to analytically tractable equations that reveal how frequency-encoded signals are systematically transformed into  distinct high-pass and low-pass responses.

In the Wave Converter (M1) module, we established a detailed analysis of the reaction kinetics (Supplementary Note 1) to characterize the system's steady-state behavior.  Under periodic light stimulation, the light-sensitive adenylyl cyclase and phosphodiesterase in M1 regulate the intracellular cAMP concentration through its synthesis (rate $k$) and hydrolysis (rate $\gamma$), respectively. The system achieves a stable oscillatory state, exhibiting a periodic waveform as demonstrated in Fig. 2C. We derived an analytical solution for this dynamic equilibrium within one period, expressed as Equation \ref{Eq 1}:  
\begin{equation}
	s(\tau) =
	\begin{cases}
		1-(1-s_\text{L})e^{-\tau},  &  0\le \tau\le \phi D  \\
		s_{\text{H}}e^{-\tau +\phi D},  &  \phi D< \tau\le \phi  \\
	\end{cases}
	\label{Eq 1}
\end{equation}
All variables and parameters are presented in non-dimensionalized form (Supplementary Table 4). The duty cycle $D$ represents the fraction of the period during which the light stimulus is active. The non-dimensional concentration $s(\tau)$ represents the relative cAMP level normalized to its theoretical maximum concentration (${k}/{\gamma}$).  The time and period are normalized to the characteristic time scale of cAMP hydrolysis ($\gamma^{-1}$),   yielding the non-dimensional time $\tau = \gamma t$ and period $\phi = \gamma T$, where $\gamma$ is the hydrolysis rate of cAMP.  Under these steady-state conditions, the non-dimensional  maximum and minimum  levels of cAMP ($s_\text{H}$ and $s_\text{L}$) are given by Equations \ref{Eq 2} and \ref{Eq 3}, respectively:  
\begin{equation}
	s_\text{H}(\phi,D)=\frac{1-e^{-\phi D}}{1-e^{-\phi }}
	\label{Eq 2}
\end{equation}
\begin{equation}
	s_\text{L}(\phi,D)=\frac{e^{\phi D}-1}{e^{\phi }-1}
	\label{Eq 3}
\end{equation}

The Thresholding Filter (M2) processes the output signal from M1 through two sequential Hill-type binding interactions: first, the cooperative binding between cAMP and the transcription factor Vfr, characterized by the microscopic dissociation constant $K_1$ (\SI{}{\micro M}), and second, the binding of the Vfr-cAMP complex to regulatory promoters, characterized by the microscopic dissociation constant $K_2$ (\SI{}{\micro M}).  These Hill processes exhibit the fastest dynamics among the three modules, with their characteristic time $T_{c2}$ being significantly shorter than $T_{c1}$. This pronounced time-scale separation allows us to assume that the dynamic response of M2 is effectively instantaneous relative to the cAMP oscillations generated by M1. Consequently, in our theoretical model,  the temporal evolution of M2's output depends solely on the time-varying input $s(\tau)$ from M1, with the two Hill processes modulating the signal amplitude (Supplementary Note  2).  Following this processing, the fraction of activated promoters $\psi(\tau)$ can be analytically expressed as Equation \ref{Eq 4}:
\begin{equation}
	\psi(\tau)=\frac{\lambda\alpha^2s(\tau)^2}{1+(\lambda+1)\alpha^2s(\tau)^2}
	\label{Eq 4}
\end{equation}
Here, $\psi(\tau)$ represents the fraction of promoters bound by Vfr-cAMP complexes relative to the total promoter population. The dimensionless parameter $\lambda = [\text{Vfr}]_0/K_2$ denotes the relative abundance of transcription factor Vfr normalized to the microscopic dissociation constant $K_2$ for Vfr-promoter binding, capturing the impact of transcription factor availability in the system. The parameter $\alpha = (k/\gamma)/K_1$ characterizes the cAMP signal strength, defined as the ratio of maximum achievable cAMP concentration ($k/\gamma$) to the microscopic dissociation constant $K_1$ for cAMP-Vfr binding, reflecting the relative strength of the cAMP signaling pathway.

The Integrator (M3) functions to integrate and average the output from M2, representing the protein expression level from promoters activated by Vfr-cAMP complexes. Assuming negligible basal expression from the regulated promoters, we derived the theoretical relationship between periodic input signals and steady-state protein expression (Supplementary Note 3). At steady state, the time-averaged protein expression level over one period can be expressed as Equation \ref{Eq 5}:
\begin{equation}
	\bar{y}=\frac{1}{\phi}\int_{0}^{\phi}\psi(\tau)\text{d}\tau
	\label{Eq 5}
\end{equation}
where $\bar{y}$ represents the dimensionless protein expression level normalized to its maximum achievable value. Through mathematical analysis of this temporal integration, we obtained an analytical solution for the steady-state expression level:  
\begin{equation}
	\begin{split}	
	\bar{y}(\alpha,\phi,D,\lambda) = & \cfrac{1}{\phi}[\frac{\lambda\alpha^2}{1+(\lambda+1)\alpha^2}{\text {ln}}(\frac{\sqrt {1+(\lambda+1)\alpha^2 s_\text{H}^2}}{\sqrt {1+(\lambda+1)\alpha^2 s_\text{L}^2}})-\\
	&\frac{\alpha \lambda/ \sqrt {\lambda+1}}{1+(\lambda+1)\alpha^2} {\text {tan}^{-1}} (\frac{\alpha\sqrt {\lambda+1}(s_\text{H}-s_\text{L})}{1+(\lambda+1)\alpha^2 s_\text{L} s_\text{H}})]+\frac{\lambda\alpha ^2 D}{1+(\lambda +1)\alpha^2}
\end{split}
	\label{Eq 6}
\end{equation}
This solution quantitatively describes how the system transforms frequency-encoded inputs into defined protein expression levels through the sequential processing of the three modules.  

In the theoretical analysis, we first examined the system behavior when the duty cycle $D$ equals 1. Under this condition, $s_\text{H}= s_\text{L}=1$, and Equation \ref{Eq 6} reduces to a pure amplitude-dependent expression:  
\begin{equation}
	y^*=\bar{y}(D=1) =\bar{y}(\alpha,\lambda)=\frac{\lambda\alpha ^2 }{1+(\lambda +1)\alpha^2}
	\label{Eq 7}
\end{equation}
where $y^*$ represents the TRGC response at full duty cycle, determined by the parameters $\alpha$ and $\lambda$.   This simplified case serves as a reference point for understanding the system's basic amplitude response characteristics.  

To analyze the frequency-dependent behavior, we identified a critical threshold in M2's filtering characteristics by examining the inflection point of Equation \ref{Eq 4} (where $\text{d}^2\psi/\text{d}s^2 = 0$). This point represents where the system's response sensitivity is maximal, as it marks the steepest rate of change in the input-output relationship. This analysis yielded a threshold value $s^*$ for the non-dimensional cAMP concentration: $s^*={1}/{\sqrt{3(\lambda+1)}\alpha}$ (Supplementary Note 4).

This threshold characterizes the filtering properties of M2, representing the concentration at which the rate of change in promoter activation is maximal. Using this threshold definition, we combined Equations \ref{Eq 6} and \ref{Eq 7} to obtain a comprehensive expression for the system response:
\begin{equation}
	\bar{y}(\alpha,f,D,\lambda) = {y^*}\ (D+G)
	\label{Eq 8}
\end{equation}
where,
\begin{equation}
	G(f,D,s^*)=f\ [\text{ln}(\sqrt{\frac{1+(s_\text{H}/\sqrt{3}s^*)^2}{1+(s_\text{L}/\sqrt{3}s^*)^2}})-\sqrt{3}s^*({\text{\text{tan}}}^{-1}(\frac{\sqrt{3}s^*(s_\text{H}-s_\text{L})}{(\sqrt{3}s^*)^2+s_\text{H}s_\text{L}}))]
	\label{Eq 9}
\end{equation}
Here, $f={1}/{\phi}$ represents the non-dimensionalized frequency.  

This formulation reveals a sophisticated decomposition of the system's response into three fundamental components (Figure 2a). First, the amplitude modulation component $y^*$ establishes the baseline response level, which is primarily governed by the system's intrinsic biochemical parameters including the relative abundance of transcription factors ($\lambda$) and the strength of cAMP signaling ($\alpha$). Second, the pulse width modulation, represented by $D$, directly captures the temporal characteristics of the input signal through its duty cycle, reflecting the proportion of time the system is actively stimulated within each period. Finally, the frequency modulation component $G$ introduces a dynamic, frequency-dependent modification to the response, enabling the system to discriminate between signals of different frequencies while maintaining the same duty cycle and amplitude. Together, these three components form a comprehensive framework that describes how the TRGC integrates and processes complex temporal signals into defined gene expression patterns.

To facilitate a more intuitive analysis of the system's behavior, we normalized the output by introducing 
\begin{equation}
	Y=\dfrac{\bar{y}}{y^*}= D+G
	\label{Eq 10}
\end{equation}
, which represents the total normalized response combining both duty cycle and frequency-dependent effects (Supplementary Note 5). This normalization allows us to examine the frequency response characteristics independently of the amplitude scaling factor $y^*$.  

To establish a comprehensive macroscopic perspective of the TRGC function, we focused on the difference between high-frequency and low-frequency responses, introducing the metric $Y_{\text{HF}}-Y_{\text{LF}}$, where $Y_{\text{HF}}$ and $Y_{\text{LF}}$ represent the normalized outputs at high and low frequencies, respectively. While the analytical expression for this difference is complex in its general form, we found that it simplifies remarkably in the limiting cases of very high and very low frequencies. The simplified expression primarily depends on two key parameters: the threshold $s^*$ and the duty cycle $D$.  

This insight led us to construct phase diagrams (Figure 2b) visualizing $Y_{\text{HF}}-Y_{\text{LF}}$ as a function of these two critical parameters. The diagrams reveal a clear dichotomy in the system's behavior. As $s^*$ increases, $Y_{\text{HF}}-Y_{\text{LF}}$ transitions from positive to negative values, indicating a shift from high-pass to low-pass characteristics. Conversely, increasing the duty cycle $D$ promotes a transition from low-pass to high-pass behavior. This dual dependence is particularly significant because $D$ serves as an experimentally controllable parameter, offering a practical means to modulate the system's frequency response within an appropriate range of $s^*$.

Through careful mathematical analysis, we identified a critical boundary between these two regimes, expressed by the relationship $D = 3{s^*}^2$. This elegant relationship (shown as dashed lines in Figure 2b; Supplementary Note 6) provides a clear demarcation between high-pass and low-pass behaviors, offering valuable guidance for circuit design and optimization.

Further quantitative analysis of the $Y_{\text{HF}}$ and $Y_{\text{LF}}$ metric revealed fundamental differences between high-pass and low-pass configurations. High-Pass FACs could achieve larger differences in response between high and low frequencies ( $\left|Y_{\text{HF}}-Y_{\text{LF}}\right|$ ) when optimally configured, providing greater potential for frequency discrimination. This enhanced capacity for distinguishing frequency differences makes high-pass configurations particularly attractive for engineering precise frequency-dependent responses. Moreover, we observed a critical limitation in low-pass configurations: for any given threshold $s^*$, low-pass responses generally produced smaller $Y$ values across their operating range. This characteristic poses practical challenges, as smaller output signals are inherently more susceptible to experimental noise and cellular stochasticity, potentially compromising measurement accuracy and reliability (Figure 2c). This inherent limitation of low-pass configurations, combined with their reduced frequency discrimination capability, provided a clear rationale for our experimental strategy. Consequently, we prioritized the investigation and implementation of High-Pass FACs, while conducting more limited studies of low-pass configurations primarily to validate our theoretical framework.

To develop a more comprehensive understanding of the system's behavior, we expanded our analysis through additional phase diagrams that explored the interplay between $s^*$, $D$, and the non-dimensionalized frequency $f$. These diagrams mapped their influences on both the frequency response function $G$  and the final output $\bar{y}(f,D,s^*)$ (Supplementary Fig. 2). This expanded parameter space exploration not only validated our theoretical predictions but also provided practical insights for optimizing circuit performance across different operating conditions.

To validate our theoretical framework, we conducted comprehensive comparisons between the CRN simulations and analytical predictions across diverse parameter spaces. The remarkable agreement between these two independent approaches demonstrates the robustness of our frequency-amplitude conversion framework in Time-Resolved Gene Circuits (TRGCs).  

The dynamic behavior of the Wave Converter (M1) reveals sophisticated signal processing capabilities (Fig. 2d). At constant periods, higher duty cycles D lead to increased maximum ($s_\text{H}$) and minimum ($s_\text{L}$) signal levels, demonstrating how the system accumulates signal during the \textquotesingle on\textquotesingle\ phase of each cycle. When duty cycle is fixed, increasing frequency causes $s_\text{H}$ to decrease while $s_\text{L}$ rises, both asymptotically approaching the duty cycle value.  This convergence of signal bounds at high frequencies reflects a fundamental characteristic of the system's temporal signal processing capability.

The Thresholding Filter (M2) exhibits sophisticated signal processing behavior that transcends conventional binary switching mechanisms (Fig. 2e). While our initial conceptual framework suggested a sharp threshold (Fig. 1b), the implemented biological system reveals a more nuanced response landscape. Rather than enforcing an abrupt transition at a fixed threshold value, M2 creates a continuous activation profile where promoter activity undergoes a gradual transition between inactive and active states. This analog processing capability emerges from the cooperative binding dynamics between cAMP and the Vfr transcription factor, with the activation threshold manifesting as a responsive range rather than a discrete point (Supplementary Fig. 1). The system's filtering characteristics can be precisely tuned through transcription factor abundance ($\lambda$), where higher $\lambda$ values systematically shift the activation profile toward lower cAMP concentrations. This tunable analog filtering mechanism not only provides more sophisticated control over frequency response characteristics but also better reflects the inherent complexity of biological signal processing.

The integration of M1 and M2 dynamics produces distinct frequency-dependent behaviors in the circuit output. Under low-pass configuration (Fig. 2f), increasing frequency progressively attenuates the expression level at fixed duty cycles. Conversely, high-pass configurations (Fig. 2f) show enhanced expression at higher frequencies, demonstrating the circuit's ability to selectively respond to different frequency ranges. This frequency selectivity emerges from the nonlinear interaction between the Wave Converter's signal processing and the Thresholding Filter's activation dynamics. 
 
To rigorously assess the predictive power of our theoretical framework, we compared normalized output predictions from both CRN simulations and analytical solutions across the entire accessible parameter space (Fig. 2g). The exceptional correlation ($R^2 = 0.992$) between these independent approaches validates the analytical model's ability to capture the essential dynamics of the system while maintaining computational efficiency. This agreement spans multiple orders of magnitude in key parameters, confirming the model's utility for both mechanistic understanding and circuit design.

These results establish our theoretical framework as a powerful tool for understanding and engineering frequency-responsive gene circuits. The framework's ability to accurately predict complex behaviors while maintaining analytical tractability provides a foundation for rational design of sophisticated genetic programs that exploit frequency-based information processing.

\subsection*{Design and Implementation of Automated Experimental Platform for TRGC Validation}
The TRGC architecture enables precise control of gene expression through both molecular and operational parameters. At the molecular level, the circuit can be tuned through CpdA and Vfr expression levels, which affect parameters $\alpha$ and $\lambda$ (Supplementary Table 5), while at the operational level, it responds to experimental parameters including light intensity ($I$) and duty cycle ($D$). These complementary control mechanisms collectively shape the circuit's output ($Y$), providing multiple degrees of freedom for engineering desired frequency responses. To systematically explore this multidimensional parameter space, we constructed 65 distinct TRGC variants with different combinations of CpdA and Vfr expression levels (Supplementary Note 7 and 8, Supplementary Fig. 3).

However, comprehensive characterization of these circuits presents significant technical challenges\cite{kwon2024advancing}. Rigorous validation requires testing multiple light intensities, periods, and duty cycles while maintaining precise control over cellular physiological states. The scale of experimentation required - thousands of experiments generating approximately 100,000 data points across 20 strains - renders traditional manual approaches impractical. Furthermore, maintaining consistent bacterial physiological states is crucial, as growth rate variations affect intracellular concentrations of key components like Vfr and CpdA. Such fluctuations would compromise experimental control parameters $\alpha$, $\lambda$ and $s^*$  (Supplementary Table 5), while growth rate-dependent changes in fluorescent protein levels could introduce measurement artifacts.

To address these challenges, we developed a high-throughput automated experimental platform capable of maintaining stable bacterial states through continuous culture (Fig. 3a). The platform integrates four core functional modules: (1) an optoplate for programmable light signal control, (2) bacterial culture agitation, (3) automated solution handling for continuous dilution, and (4) fluorescence measurement. This integrated system enables parallel testing of n × 96 experimental samples while providing independent control of light signal parameters for individual wells (Supplementary Fig. 4). The continuous dilution culture maintains stable growth rates, preventing protein concentration fluctuations from growth phase transitions, while automated fluorescence measurements enable systematic data collection.

The automated workflow follows a rigorously controlled process (Fig. 3b). Under constant temperature and agitation conditions, bacterial strains receive independently programmed light signals via the OPCU (optoplate control unit) device. Hourly sampling cycles remove \SI{50}{\micro \liter}  for fluorescence measurement while the remaining culture undergoes rapid dilution (4 minutes) with fresh medium. This precise temporal control ensures cultivation continuity while minimizing perturbations to bacterial growth states (Supplementary Note 9, Supplementary Fig. 5).

Platform validation demonstrated exceptional stability and reproducibility. The system maintained optical density at $0.09\pm0.01$ across 96 parallel samples over extended periods ($>12\ \text{h}$) (Fig. 3c). Even with varying initial conditions, continuous dilution established consistent optical densities within $\sim4$ hours. Cross-batch reproducibility analysis revealed excellent consistency in fluorescence measurements, with coefficients of variation below 10\% (Fig. 3d).

To systematically characterize circuit behavior, we first mapped the relationship between key parameters ($\gamma$ and $s^*$) and frequency response characteristics (Fig. 3e). The resulting phase diagram reveals distinct regions corresponding to different frequency-amplitude conversion behaviors, with $Y_{\text{HF}}-Y_{\text{LF}}$ serving as a metric for frequency discrimination capability. For this metric, $Y_{\text{LF}}$ was evaluated at frequencies corresponding to bacterial division cycles (approximately $1/2400\ \text{s}^{-1}$), ensuring stable cellular states during measurement periods. The high-frequency response ($Y_{\text{HF}}$) was assessed at $1/100\ \text{s}^{-1}$, establishing an experimentally accessible range that respects cellular physiological constraints. We strategically selected 29 representative strains from our engineered TRGC variants, shown as circular markers in the phase diagram, to systematically sample different regions of the theoretical parameter space. These strains were specifically chosen to validate our theoretical predictions across diverse operating regimes while maintaining experimental feasibility within biological constraints.

Comprehensive frequency response characterization using our automated platform revealed remarkable agreement between theoretical predictions and experimental measurements (Supplementary Note 10). Figure 3f presents frequency response curves for all strains at fixed duty cycle $D = 0.3$, demonstrating consistent alignment across diverse parameter combinations and validating our theoretical framework's predictive power (Supplementary Note 11).

Our theoretical analysis (Figure 2b) identified duty cycle ($D$) as a crucial parameter in modulating frequency responses. We validated this prediction through two complementary experimental approaches. First, we characterized several TRGC variants at $D = 0.1$ (Fig. 3g), demonstrating that circuits maintain their frequency discrimination capabilities even at low duty cycles while exhibiting systematic variations in response amplitude and frequency sensitivity based on molecular parameters. Second, detailed characterization of a representative strain (FAC03C17V17) across multiple duty cycles (Fig. 3h) revealed how duty cycle modulation systematically alters both response magnitude and frequency sensitivity while maintaining fundamental high-pass characteristics.

A particularly significant finding emerged from strain FAC03C22V34, which demonstrated controlled switching between High-Pass and Low-Pass FAC behaviors through precise manipulation of light intensity and duty cycle (Fig. 3i). This dynamic control over frequency response characteristics revealed the circuit's programmable nature. By mapping these behavioral transitions onto the theoretical phase diagram (Fig. 3j), where solid circular points and connecting arrows track parameter-driven changes, we provided direct experimental validation of the predicted phase boundaries between distinct operating modes.

To rigorously validate our theoretical framework, we conducted a comprehensive analysis comparing experimental measurements with theoretical predictions across our entire strain library under diverse conditions (Fig. 3k). The strong correlation between predicted and observed normalized outputs ($Y$) spans multiple strains, duty cycles, and frequencies, demonstrating the broad applicability and robust predictive power of our model across the accessible parameter space.

These detailed characterizations establish the TRGC architecture as a versatile platform for implementing programmable frequency-dependent gene regulation. Our automated experimental system enabled systematic exploration and verification of complex dynamic behaviors across multiple strains and conditions, providing a strong foundation for engineering sophisticated frequency-responsive gene circuits.

\subsection*{Frequency Signal Control Expands Multi-gene Expression State Combinations}
Having established the FAC's ability to process frequency-encoded signals, we next investigated whether this additional control dimension could expand the capabilities of gene regulation. While traditional amplitude-based control relies solely on signal intensity, frequency modulation introduces new possibilities for fine-tuned gene expression. We hypothesized that in a one-to-many regulatory architecture, where multiple genes respond to a single input signal, frequency-based control could access expression states unreachable through amplitude modulation alone.

To test this hypothesis, we first examined a two-protein system where genes exhibit different sensitivities to Vfr-cAMP complex regulation, characterized by distinct dissociation constants  $K_2$ and corresponding $\lambda$ values ($\lambda_{\text{A}}$ and $\lambda_{\text{B}}$) in Equation 6. This configuration creates a two-dimensional state space ($Y_{\text{A}}$, $Y_{\text{B}}$) of protein expression (Equation 10). While traditional approaches rely on light intensity ($I$) and duty cycle ($D$) as control parameters, our TRGC circuit introduces frequency ($f$) as an additional control dimension, potentially expanding the accessible gene expression state combinations (Figure 4a).

To quantitatively analyze state space accessibility, we introduced a discretization parameter  $\epsilon = 0.1$, which functions as a measure of resolution in state space partitioning. Similar to how the minimum number of spheres $N(R)$ needed to cover a point set scales inversely with sphere radius $R$ as $N(R) \sim 1/R^d$\cite{gneiting2012estimators}, the total number of theoretically distinguishable states scales inversely with our discretization parameter $\epsilon$. This parameter divides each dimension of normalized expression ($Y$) into $1/\epsilon$ equal intervals. In a two-promoter system, our choice of $\epsilon$ creates a $10\times10$ grids with $100$ theoretically possible distinct expression states (Figure 4b). While this discretization provides a simplified metric for quantifying state space expansion, it's important to note that the actual resolution of distinguishable gene expression states is fundamentally limited by intrinsic cellular noise. Therefore, this grid-based quantification serves primarily as a qualitative measure to demonstrate relative changes in state space accessibility\cite{kaern2005stochasticity,mar2009decomposition}. 

Under pure amplitude modulation through varying light intensity, we observed only 19 distinct states, represented by the blue curve and corresponding grid cells in Figure 4b. However, introducing frequency modulation revealed a remarkable expansion of accessible states. As frequency was varied from $1$ to $1 \times 10^{-3}$ (non-dimensionalized units), the system accessed new regions of state space (red grid cells), increasing the total number of unique state combinations from 19 to 38 (Figure 4b). This state space expansion demonstrates a fundamental capability of frequency-based regulation that exceeds traditional amplitude control. Notably, our theoretical analysis predicts that as frequency approaches zero, $Y_{\text{A}}=Y_{\text{B}}=D$, creating a diagonal line in state space as duty cycle varies (Supplementary Note 12). This limiting behavior provides a theoretical boundary for the system's operational range.

We extended this analysis to three-gene systems, maintaining $\epsilon = 0.1$ while exploring the expanded state space ($Y_{\text{A}}$,  $Y_{\text{B}}$, $Y_{\text{C}}$). To visualize this three-dimensional state space, we mapped the expression levels of the three genes to a color space, where different colors represent distinct combinations of expression intensities (Figure 4c). The impact of frequency modulation proved even more pronounced in this higher-dimensional space, increasing accessible states from 27 under amplitude modulation to 95 distinct combinations under frequency control.

To experimentally validate these theoretical predictions, we conducted high-throughput screening of 260 promoter candidates to identify sets with appropriate $\lambda$ values (Supplementary Note 13, Supplementary Fig. 7). We constructed both two-gene systems expressing sfGFP and CyOFP (Figure 4d) and three-gene systems expressing sfGFP, CyOFP, and mScarlet (Figure 4e). Using the same color mapping scheme as in our theoretical analysis, the experimental results demonstrated clear state space expansion through frequency modulation in both two- and three-dimensional cases, confirming our theoretical predictions (Supplementary Note 14).

Our theoretical analysis based on Equation 6 demonstrates that the capacity for expanding the repertoire of gene expression states scales proportionally with the regulatory dimension (Supplementary Note 12, Supplementary Fig. 6). This mathematical prediction is supported by our experimental observations: while two-gene systems showed a 2-fold increase in accessible states through frequency modulation, three-gene systems achieved a 3.5-fold expansion, from 27 to 95 distinct states. This systematic relationship between regulated gene number and state space expansion suggests frequency modulation becomes increasingly advantageous for coordinating larger gene networks, offering a scalable approach to expand the programmable range of cellular states beyond traditional amplitude-based control.

These findings have profound implications for synthetic biology and cellular engineering. The ability to access an expanded state space through frequency modulation provides new opportunities for engineering sophisticated cellular behaviors. By enabling more nuanced control over multiple genes simultaneously, this approach opens new possibilities for designing complex biological circuits with enhanced computational and regulatory capabilities.

\section*{Summary and Perspective}
Our work establishes a comprehensive framework for implementing and understanding frequency modulation in synthetic gene circuits, addressing a critical gap between natural and engineered biological systems\cite{lu2009next}. Through the development of Time-Resolved Gene Circuit (TRGC) architecture, we demonstrate how cells can effectively process frequency-encoded information, expanding the traditional amplitude-modulation paradigm in synthetic biology.

 The emergence of dynamic signal processing in gene circuit design marks a significant advancement in cellular information processing\cite{purvis2013encoding}. Beyond conventional amplitude-based regulation, periodic signals introduce additional control dimensions through frequency and duty cycle modulation. These can be categorized into two distinct modes: pulse-width modulation (PWM)\cite{levine2013functional}, where signal duration varies within a period, and pure frequency modulation, where duty cycle remains constant while frequency varies. While PWM represents an important advancement in dynamic cellular control\cite{chuang2014synthesizing,benzinger2018pulsatile,benzinger2022synthetic,ravindran2022synthetic}, it fundamentally modulates the time-averaged signal amplitude through varying pulse widths, making it an extension of amplitude-based regulation rather than true frequency modulation. In contrast, pure frequency modulation maintains constant duty cycle and thus constant time-averaged signal intensity\cite{tomazou2018computational}, enabling information encoding exclusively in the frequency domain. This distinction is particularly significant as it establishes frequency as an independent control dimension, separate from amplitude-based mechanisms.   

Our approach integrates control theory principles with synthetic biology to create a robust theoretical framework. The Frequency-Amplitude Converter (FAC) architecture demonstrates how biological systems can transform frequency-encoded signals into amplitude-modulated gene expression patterns, analogous to engineered FM systems. This framework, combining Chemical Reaction Network\cite{van2015programmable} simulations with theoretical analysis\cite{kaern2005stochasticity}, provides a multi-scale understanding from molecular interactions to system-level behaviors. The successful experimental validation through our automated platform demonstrates the feasibility of implementing sophisticated control strategies in living cells.

The development of our automated experimental platform represents a significant advancement in synthetic biology's Design-Build-Test-Learn (DBTL) cycle\cite{appleton2017design}. By enabling high-throughput testing while maintaining stable physiological conditions, this platform facilitates rigorous validation of theoretical predictions that would be impractical through manual experimentation. This integration of theory and automated experimentation establishes a new paradigm for studying complex dynamic behaviors in synthetic biology.

Our experimental findings, coupled with observations from natural systems, provide compelling evidence for the significance of frequency-modulated signal processing in cellular systems. We demonstrated that frequency modulation expands signaling pathway capabilities by enabling global regulation of multiple target genes with different activation thresholds or affinities. This was particularly evident in our multi-gene system experiments, where frequency modulation accessed an expanded state space of gene expression combinations. This capability mirrors natural systems, where frequency modulation facilitates proportional co-regulation of diverse target genes, as observed in the pulsatile behavior of transcription factors such as \textit{p}53\cite{purvis2012p53}, \textit{Ascl}1\cite{imayoshi2013oscillatory}, \textit{NF$\kappa$B}\cite{tay2010single}, and the SOS stress response system\cite{locke2011stochastic}. Notably, frequency modulation provides a mechanism linking individual protein dynamics to large regulon expression, suggesting its role in orchestrating genome-scale expression patterns. Given its observed functions in protein and metabolic networks, as well as transcriptional regulation, we anticipate that frequency-modulated regulation may represent a general principle by which cells process and respond to dynamic environmental signals.

From a synthetic biology perspective, frequency-modulated circuits offer distinct advantages through their relatively simple genetic architecture\cite{lu2009next}. The FAC system requires a modest set of genetic elements compared to traditional synthetic circuits that often demand multiple regulatory components and precise expression balancing. This architectural simplicity, combined with sophisticated control capabilities, suggests that integrating frequency-modulated regulation into synthetic circuits is not only feasible but potentially transformative. Moreover, incorporating dynamic frequency-based control into engineered circuits presents unique opportunities to effectively address and exploit inherent cellular characteristics, such as noise management and shared regulatory resource allocation\cite{eldar2010functional}. This approach to circuit design opens new possibilities for engineering cellular behaviors in ways previously unexplored in traditional engineering disciplines.

Looking forward, this understanding of frequency-based signal processing opens new avenues in both fundamental research and practical applications\cite{li2018frequency}. In metabolic engineering, the coordination of multiple genes through frequency modulation could enable sophisticated pathway control. The expanded state space accessible through frequency modulation provides new tools for fine-tuning cellular behavior and controlling complex phenotypes. Furthermore, the connection between single-protein dynamics and genome-wide expression patterns may offer insights into coordinated cellular responses across different organizational scales, potentially revealing new principles for both synthetic biology design and natural regulatory networks.

Several important challenges and opportunities remain for future research. These include developing more sophisticated frequency-responsive elements, improving methods for temporal control of cellular systems, and better understanding noise characteristics in frequency-modulated circuits. Additionally, exploring frequency modulation in diverse cellular contexts and organisms could reveal new applications and design principles, further expanding the potential of this regulatory approach in synthetic biology.

\section*{Methods}
%\subsection*{Probe Performance Characterization}
%The experimental approach for probe performance characterization is consistent with that used in the development of the green cAMP probe\cite{wang2022high}.
\subsection*{Computational Model}
In this study, we employ deterministic mass-action chemical reaction network (CRN) models to simulate the system. The genetic circuit TRGC, comprising 10 distinct species and 9 chemical reactions (denoted as $r1$ to $r9$ in Supplementary Table 1), is at the core of these models. Illustrated in Supplementary Fig. 1, the diagram provides a more realistic representation of the CRN within a bacterium. The reaction parameters used in this model, drawn from existing literature or our experimental data, are detailed in Supplementary Table 3. In this model, we simplify the processes of transcription and translation of CpdA and Vfr, treating them as constants determined by the bacterial strain. The simulation model further streamlines the transcription and translation of proteins into a single step, while introducing a deactivation stage for bPAC to incorporate the non-instantaneous deactivation of bPAC post light exposure.

\subsection*{Cultivation of Bacterial Strains}
In this study, genetically engineered \textit{Pseudomonas aeruginosa} strains were cultured at \SI{37}{\degreeCelsius}. Strains carrying the bPAC fragment needed to avoid light throughout the cultivation process. The detailed cultivation protocol involved streaking the strains stored at \SI{-80}{\degreeCelsius} on classical LB agar plates shielded from light with foil. After overnight incubation for resuscitation, single colonies were selected and transferred to FAB culture medium\cite{xiong2024optimal} containing \SI{30}{m M} glutamate and \SI{1}{\micro M} $\text{FeCl}_3$ . The cultures were agitated at 220 rpm until the $\text{OD}_{600}$ measurement reached around 0.5. The concentrations of antibiotics used during bacterial cultivation are \SI{30}{\micro\gram/\milli\liter}  gentamicin, \SI{100}{\micro\gram/\milli\liter} tetracycline, and \SI{150}{\micro\gram/\milli\liter} carbenicillin. 

\subsection*{Construction of Bacterial Strains}
All plasmids and strains are listed in Supplementary Table 6-7.  Unless otherwise specified, the knockout of all genes and seamless insertion of gene fragments into the genome in this study were accomplished using CRISPR technology. The construction of relevant plasmids was carried out using Gibson assembly. Refer to Supplementary Note 7 for additional details on the construction strategies for more bacterial strains and plasmids. The chassis strain FAC01: PAO1-$\Delta$\textit{pslBCD}$\Delta$\textit{pelA}$\Delta$\textit{exoS}$\Delta$\textit{exoT}$\Delta$\textit{cyaA}$\Delta$\textit{cyaB} was constructed by successive knockout of six gene clusters. The experimental procedure was refined based on existing literature\cite{chen2018crispr}, with the deletion of the \textit{cyaA} gene as an example. The specific experimental procedure is as follows: (1) Construct a plasmid PCRISPR-\textit{cyaA} containing the gRNA and homologous recombination segment. (2) Transform the plasmid PCASPA containing Cas9 into the PAO1 strain, and electroporate the plasmid PCRISPR-\textit{cyaA}, and plate on a double-resistant plate containing tetracycline and carbenicillin. (3) PCR confirm the successful knockout of the target gene \textit{cyaA} in the resulting colonies. Pick colonies and culture overnight on LB agar plates without sodium chloride at 15\% (wt) surcose to loss plasmids. Sequence verification will confirm the PAO1-$\Delta$\textit{cyaA} strain. Subsequent knockouts of genes like \textit{pslBCD}, \textit{pelA}, \textit{exoS}, \textit{exoT} and \textit{cyaB} can be performed in a similar manner in the PAO1-$\Delta$\textit{cyaA} strain. Subsequently, through the integration of the PA1/O4/O3-\textit{bPAC} fragment into the FAC01 genome utilizing the CTX transposon insertion system, the engineered strain FAC03 was successfully generated. 

In the wild-type PAO1 strain, the expression of the \textit{vf}r and \textit{cpdA} genes is regulated by cAMP. To eliminate this specific influence, we used CRISPR technology to seamlessly replace the promoters of these genes with constitutive promoters in the FAC03 bacterial genome (Supplementary Note 8). The selectable RBS options encompass B0034-RBS046-RBS004-RBS017-RBS021\cite{fu2023programming}, while the available promoters include J23106-J23115-J23110-J23100-J23102. Through diverse combinations, a total of 65 strains were systematically engineered. 

To assess the intracellular expression levels of cAMP, we constructed a plasmid designated as Plac-\textit{sfGFP}-T0T1-J23102-\textit{CyOFP}-pJN105 and subsequently electroporated it into various chassis cells. Unless otherwise specified, all strains referenced in the main figure contain this plasmid. The constitutively expressed CyOFP fluorescent protein serves as an internal standard for normalizing bacterial growth differences. The change in intracellular cAMP concentration is calculated by comparing the ratio of sfGFP to CyOFP. 

\subsection*{Automated Experiment}
The automation island depicted in Figure 3a serves as a consolidated area where automation instruments, equipment, and control modules catering to specific experimental functions are harmoniously integrated. Components include robotic arms, Microplate Reader (Tecan, Spark), Incubator (LiCONNic, STX44-ICBT), Liquid Handler (Tecan, Fluent 780), Plate Hotel (LiCONNic, LPX220), Microplate Washer (Tecan, HydroFlex) and Self-developed Steering platform. The intelligent control system comprises various modules such as process editing, task scheduling, data analysis, equipment management, etc. For instance, the task scheduling system can precisely coordinate equipment operations to achieve automated workflow, and the data analysis system can collect and analyze experimental data in real-time. 

Supplementary movie 1 presents a comprehensive demonstration of an automated experimental workflow. The automated platform allows for concurrent parallel execution of multiple tasks. Using our self-developed light control device OPCU, we can program the input of light signal intensity $I$, period $T$, and duty cycle $D$. Subsequently, dispense the bacterial solution into black 96-well plates (LABSELECT, 11514), install the plate into the OPCU, and place it in the WareHotels of the automation island. Then, start the experiment to achieve continuous dilution of bacteria and data collection. The procedure and script for the automated experiment are detailed in Supplementary Fig. 4 and Supplementary Note 9.

\subsection*{Bacterial Image Acquisition}
After the completion of the automated experiment, following the method previously described, rapid high-throughput acquisition of microscopic images of single bacteria in a 96-well plate is conducted. The experimental steps are outlined as follows: Firstly, prepare a 1\% agar plate of FAB medium in a 96-well format, with the composition of the medium identical to that used in the automated experiment. Next, pipette \SI{6}{\micro \liter} of the bacterial suspension onto the corresponding wells of the agar plate. Finally, compress the bacterial suspension to a thickness of 0.17 mm on a specialized microscope slide.  Each well corresponds to a different bacterial strain, and the OPCU is used to set independent illumination conditions for each well. Utilize a fluorescence microscope (IX-71; Olympus) equipped with a 100× oil objective to capture four fields of view for each well, with approximately 500 bacteria per field, acquiring fluorescent images. Fluorescent images of sfGFP, CyOFP and mScarlet were acquired  by two Zyla 4.2 Scmos cameras. The fluorescence of sfGFP, CyOFP and mScarlet was excited using a solid-state light source (Lumencor Spectra X).

\subsection*{Data Analysis}
In automated experiments, we exported absorbance measurements at 600 nm and fluorescence intensity readings for sfGFP (470-520 nm), CyOFP (488-590 nm), and RFP (560-610 nm) at different time points from the plate reader. For strains with internal standards, i.e., fluorescent proteins with constitutive promoter expression, we normalized the fluorescence intensity of the fluorescent protein of interest by dividing it by the internal standard fluorescence value. For strains without internal standards, such as those with three cAMP-induced promoters fused with different fluorescent proteins, we normalized the corrected fluorescence intensity values using the absorbance value $\text{OD}_{600}$ . Then, we subtract the normalized expression level from the values obtained under continuous light exposure, and then normalize by subtracting the difference between the values obtained under continuous light exposure and those obtained for the light-avoiding group. This process yields the numerical value of the corresponding output $Y$ in the theoretical formula.  By plotting the output values $Y$ against variable frequencies and fitting them with an analytical formula, we can determine the parameters ,such as $\lambda$, $\gamma$, and others that need to be fitted in the system. Detailed data fitting procedures are provided in the Supplementary Note 11. All experiment data are expressed as the mean ± SD. 
\backmatter
\section*{Supplementary Information}
Additional results, additional tables and figures can be found in Supplementary Information. 
\section*{Acknowledgements}
Special thanks to Aiguo Xia for guidance and assistance with microscope usage, to Lei Wang for developing the experimental setup, to Shenzhen Infrastructure for Synthetic Biology for providing experimental platform and technical support for the automated characterization in this work. This work was supported by the National Key Research and Development Program of China (Grant No. 2020YFA0906900  to Fan Jin), the National Natural Science Foundation of China (Grant No. 32101177 to Yajia Huang), the National Natural Science Foundation of China (Grant No. 32101002 to Ye Li), Shenzhen Engineering Research Center of Therapeutic Synthetic Microbes (Grant No. XMHT20220104015 to Fan Jin) and the Innovation Foundation of National Science Library (Chengdu) (Grant No. E3Z0000903 to Shuai Yang).
\section*{Author Contributions}
We contributed to this study as follows: Project Conceptualization and Supervision (Fan Jin led the conceptual development; Fan Jin and Shuai Yang provided project supervision); Theoretical Framework Development (Fan Jin developed the foundational framework including Fig. 1a \& Fig. 2a and derived Equations 1 to 10; Jiarui Xiong established threshold theory, characterized threshold-duty cycle relationships, and extended analysis to frequency modulation and n-dimensional spaces; Shuai Yang and Shengjie Wan conducted theoretical validation through simulations); FAC Multi-gene Control System (Fan Jin provided conceptual framework; Rongrong Zhang designed experimental approaches; Shengjie Wan performed simulation validation); Automated Experimental Platform (Rongrong Zhang developed experimental processes; Ye Li integrated hardware and software systems); Chemical Reaction Network Analysis (Fan Jin, Shuai Yang, Shengjie Wan, Jiarui Xiong, and Lei Ni collaborated on CRN construction and analysis); Computational Studies (Shengjie Wan and Shuai Yang performed simulations); Gene Circuit Implementation (Shuai Yang developed design framework and simulations; Rongrong Zhang performed experimental characterization; Shengjie Wan conducted parameter fitting); Molecular Biology (Rongrong Zhang, Yajia Huang, Shuai Yang, Bing Li, and Mei Li performed molecular cloning and primer design); Promoter Library Development (Shuai Yang, Rongrong Zhang, and Bing Li conducted screening and characterization); Automated Experimental Development (Rongrong Zhang and Ye Li optimized automated workflows); Data Analysis (Shuai Yang, Shengjie Wan, and Lei Ni developed analysis pipelines; Shengjie Wan, Rongrong Zhang, and Shuai Yang analyzed experimental data; Shengjie Wan, Rongrong Zhang, and Jiarui Xiong performed model fitting); Strain Development (Rongrong Zhang, Bing Li, and Mei Li maintained bacterial cultures); Protocol Development (Rongrong Zhang and Shuai Yang established and validated experimental procedures); Manuscript Preparation (Shuai Yang wrote main text and contributed figures; Rongrong Zhang, Shengjie Wan, and Jiarui Xiong contributed figures, movies, and supporting information); Manuscript Review (Fan Jin, Shuai Yang, Rongrong Zhang, Shengjie Wan, and Jiarui Xiong revised and reviewed the manuscript).
\section*{Declarations}

The authors declare no competing interests.
\bibliography{Ref}
\newpage
\begin{figure}[h]
	\centering
	\includegraphics[width=1.0\textwidth]{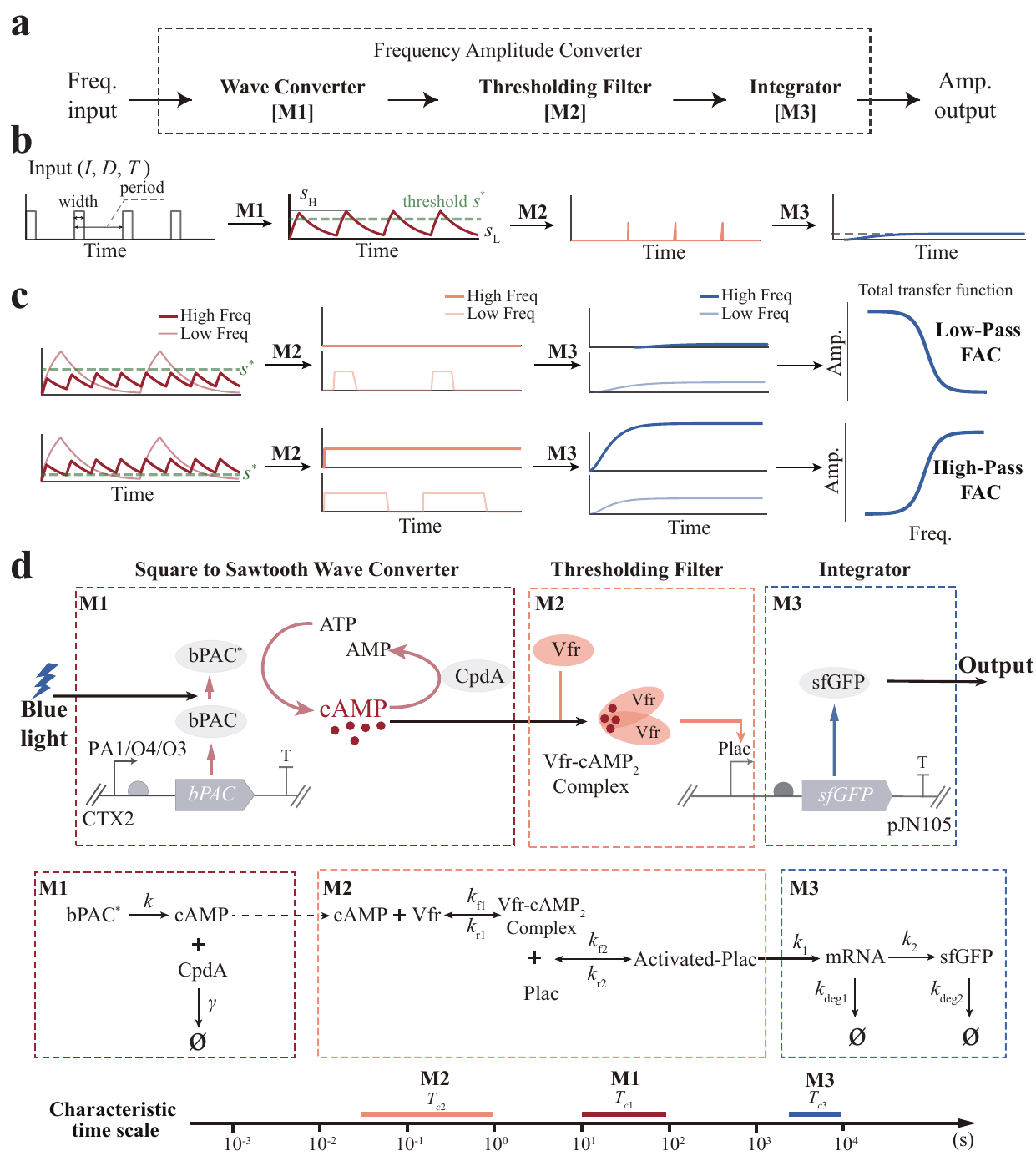}
	\raggedright
	\caption{\textbf{Frequency-Amplitude Converter (FAC) architecture and implementation in synthetic gene circuits.} \
\normalfont (a) Modular architecture of the FAC system. Three sequential modules - Wave Converter (M1), Thresholding Filter (M2), and Integrator (M3) - process input signals through distinct temporal domains for frequency-to-amplitude signal conversion. (b) Signal processing cascade through the FAC system. Periodic inputs are transformed into sawtooth waveforms by the Wave Converter, selectively processed by the threshold-based filter, and temporally integrated to generate defined output amplitudes. This sequential processing enables conversion of frequency-encoded signals into specific gene expression levels. (c) Filtering characteristics of FAC configurations determined by threshold settings ($s^*$). Upper panel: High threshold settings establish low-pass behavior through selective attenuation of high-frequency signals. Lower panel: Low threshold settings create high-pass behavior by preferential transmission of high-frequency signals. Signal processing dynamics emerge from the interaction between waveform characteristics and threshold selection. (d) Molecular implementation and temporal dynamics of the Time-Resolved Gene Circuit (TRGC). Upper panel: Optogenetic circuit design incorporating light-activated bPAC and CpdA phosphodiesterase (M1),} % cAMP-dependent Vfr transcription factor binding (M2), and protein expression machinery (M3). Lower panel: Chemical Reaction Network (CRN) analysis demonstrating temporal segregation of module dynamics. Characteristic time constants ($T_{c1}$, $T_{c2}$, $T_{c3}$) differ by orders of magnitude ($T_{c2} \ll T_{c1} \ll T_{c3}$), enabling efficient sequential processing through temporal decoupling.}
\label{fig1}		
\end{figure}
\clearpage
\begin{figure}[h]
	\raggedright
	\captionof*{figure}{\normalfont cAMP-dependent Vfr transcription factor binding (M2), and protein expression machinery (M3). Lower panel: Chemical Reaction Network (CRN) analysis demonstrating temporal segregation of module dynamics. Characteristic time constants ($T_{c1}$, $T_{c2}$, $T_{c3}$) differ by orders of magnitude ($T_{c2} \ll T_{c1} \ll T_{c3}$), enabling efficient sequential processing through temporal decoupling.}
\end{figure}

\newpage	
\begin{figure}[h]
	\centering
	\includegraphics[width=1.0\textwidth]{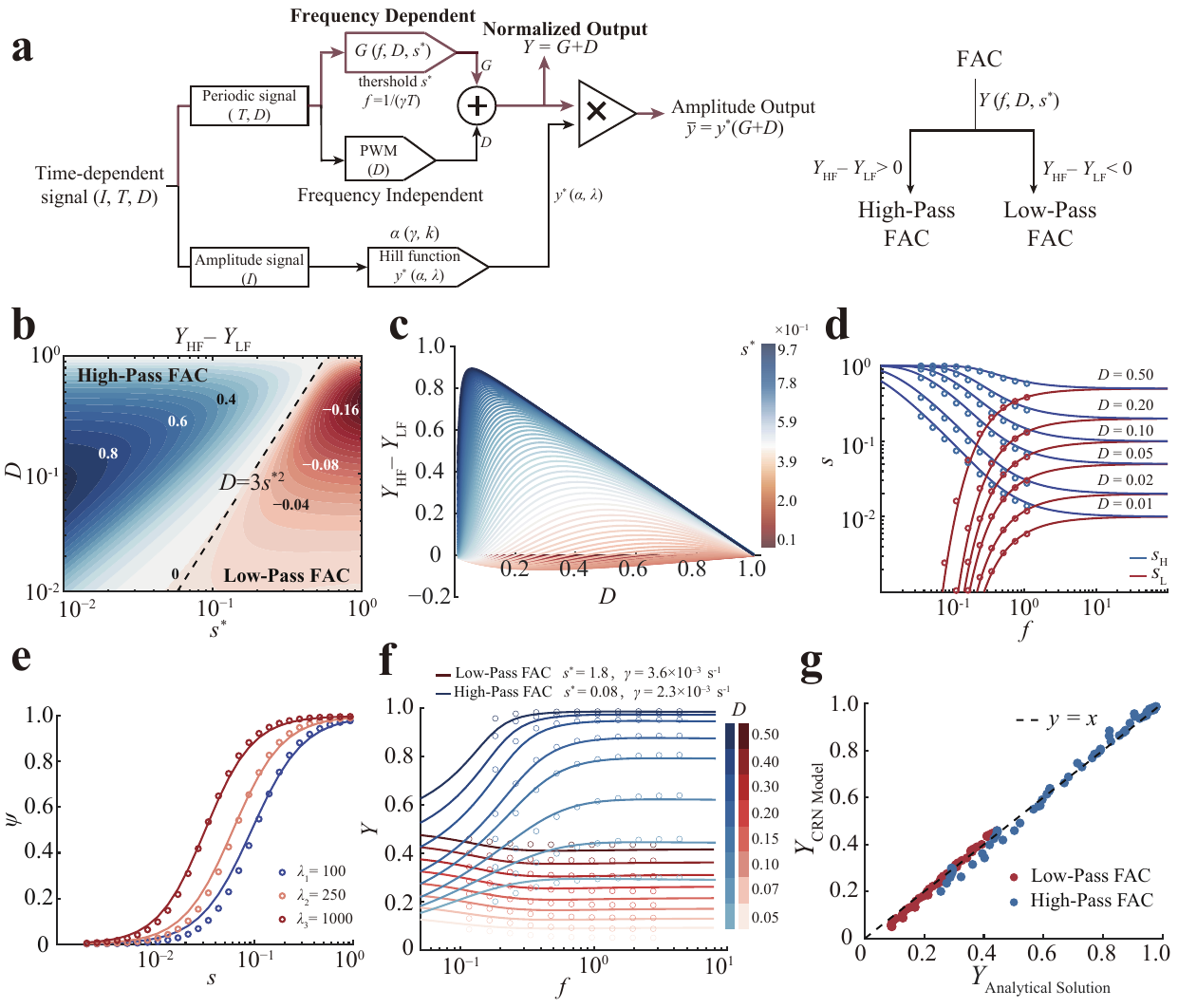}
	\raggedright
	\caption{\textbf{Theoretical framework and validation of the Time-Resolved Gene Circuit (TRGC).}\
\normalfont (a) Decomposition of the TRGC response into three fundamental components: amplitude modulation ($y^*$), pulse width modulation ($D$), and frequency modulation ($G$). Each component contributes distinctly to the overall circuit behavior. (b) Phase diagrams illustrating the transition between high-pass and low-pass frequency-Amplitude converter (FAC) behaviors. The dashed line ($D = 3{s^*}^2$) represents the theoretical boundary between these regimes. Color scale indicates the difference between high-frequency and low-frequency responses ($Y_\text{HF} > Y_\text{LF}$), where responses are evaluated at \SI{1e-1}{s^{-1}}  and \SI{1e-5}{s^{-1}}, respectively. Blue regions indicate high-pass behavior ($Y_\text{HF} > Y_\text{LF}$, while red regions represent low-pass behavior ($Y_\text{HF} < Y_\text{LF}$). (c) Quantitative analysis of frequency discrimination capability ($Y_\text{HF} - Y_\text{LF}$) as a function of duty cycle (D) for both high-pass and low-pass configurations. High-pass FACs demonstrate larger absolute differences in frequency response compared to Low-Pass FACs, indicating enhanced frequency discrimination capability. The contrasting magnitudes provide a theoretical basis for prioritizing high-pass configurations in experimental implementation. (d) Comparison of maximum ($s_\text{H}$) and minimum ($s_\text{L}$) signal levels as functions of frequency and duty cycle. Solid lines represent theoretical predictions; circles show CRN simulation results. The convergence of sH and sL at high frequencies demonstrates the fundamental limits of signal processing in the system. (e) Characterization of the Thresholding Filter's (M2) activation profile. The continuous transition between inactive and active states reveals analog processing capabilities, modulated by transcription factor abundance ($\lambda$). Inset shows the shift in activation threshold with varying $\lambda$ values. (f) Normalized output ($Y$) under low-pass and high-pass FAC configurations. Solid lines show theoretical predictions; points represent CRN simulation results across different duty cycles ($D$). The contrasting frequency responses demonstrate the circuit's ability to implement distinct filtering behaviors. (g) Correlation between theoretical predictions and CRN simulations across the complete parameter space ($R^2 = 0.992$), validating the analytical framework's accuracy and robustness. Data points represent different combinations of frequency, duty cycle, and circuit parameters.}
\label{fig2}
\end{figure}
\clearpage
\newpage

\begin{figure}[h]
\centering
	
	\includegraphics[width=1.0\textwidth]{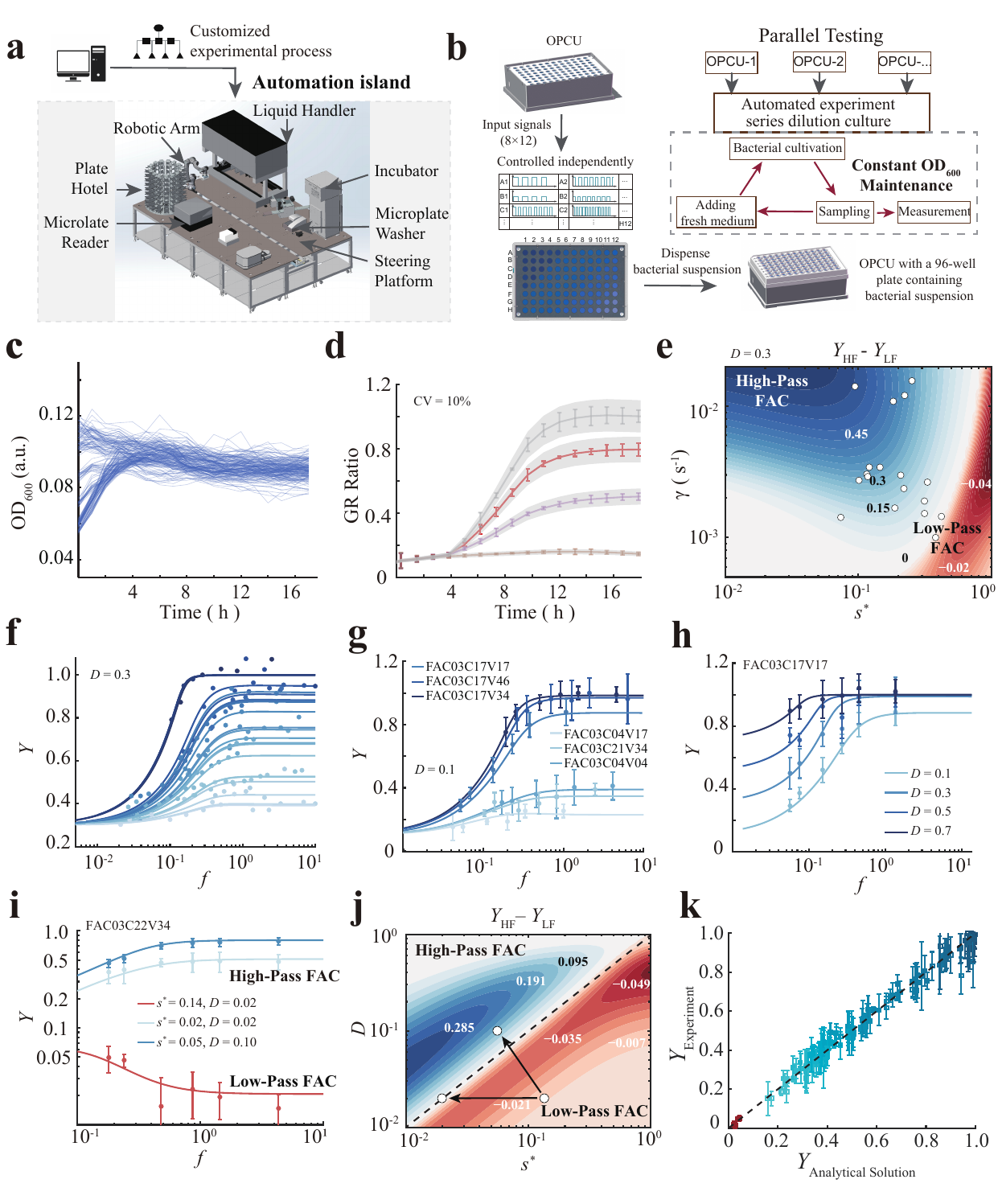}
	\raggedright
	\caption{\textbf{High-throughput automated platform enables systematic characterization of TRGC frequency responses.} \
		\normalfont (a) Schematic of the automated experimental platform integrating four core modules: optoplate signal control, bacterial culture shaking, bacterial solution dilution, and fluorescence measurement. The platform enables parallel testing of n × 96 samples while maintaining precise control over experimental conditions. (b) Workflow diagram of the automated experimental process. Under constant temperature and agitation, bacterial strains receive programmed light signals via OPCU device. Hourly cycles alternate between \SI{50}{\micro\litre} sampling for fluorescence measurement and rapid dilution (4 minutes) with fresh medium, ensuring stable growth conditions. (c) Platform stability validation demonstrating consistent control of bacterial states over extended periods ($>12\ h$). Plot shows optical density maintenance at $0.09 \pm 0.01$ across 96 parallel samples, with convergence of varying initial conditions within $\sim 4$ hours through continuous dilution. (d) Cross-batch  }%analysis showing fluorescence measurement consistency with coefficients of variation below 10\%, validating the platform's reliability for long-term experiments. (e) Phase diagram mapping the relationship between circuit parameters ($\gamma$ and $s^*$) and frequency response behavior ($Y_\text{HF} < Y_\text{LF}$) at duty cycle $D = 0.3$,where responses are evaluated at \SI{1.00e-2}{s^{-1}} and \SI{4.17e-4}{s^{-1}}, respectively. Circular markers indicate the 20 engineered TRGC variants strategically positioned to sample different regions of the theoretical parameter space. (f) Frequency response curves for all strains shown in (e) at $D = 0.3$. Solid lines represent theoretical predictions; data points show experimental measurements, demonstrating consistent agreement across diverse parameter combinations. (g) Frequency response characterization of selected TRGC variants at $D = 0.1$, showing distinct High-Pass FAC behaviors. Results validate maintenance of frequency discrimination capabilities at low duty cycles while demonstrating systematic variation in response characteristics based on molecular parameters. (h) Detailed characterization of strain FAC03C17V17 across multiple duty cycles, revealing systematic modulation of frequency response profiles. Lower duty cycles enhance frequency discrimination with reduced output amplitude, while higher duty cycles produce stronger outputs with modified frequency sensitivity.(i) Dynamic switching between high-pass and Low-Pass FAC behaviors in strain FAC03C22V34 through coordinated manipulation of light intensity and duty cycle. (j) Phase diagram mapping of transitions observed in (i), with solid circular points and arrows indicating parameter-driven switches between high-pass and low-pass regimes.Responses are evaluated at\SI{1.00e-2}{s^{-1}} and \SI{4.17e-4}{s^{-1}}. (k) Comprehensive correlation analysis between theoretical predictions and experimental measurements across the entire strain library under diverse conditions, validating the broad applicability of the theoretical framework($R^2=0.986$). Error bars represent standard deviation from multiple experimental batches.} 
\label{fig3}
\end{figure}
\clearpage
\begin{figure}[h]
	\raggedright
	\captionof*{figure}{\normalfont reproducibility analysis showing fluorescence measurement consistency with coefficients of variation below 10\%, validating the platform's reliability for long-term experiments. (e) Phase diagram mapping the relationship between circuit parameters ($\gamma$ and $s^*$) and frequency response behavior ($Y_\text{HF} < Y_\text{LF}$) at duty cycle $D = 0.3$,where responses are evaluated at \SI{1.00e-2}{s^{-1}} and \SI{4.17e-4}{s^{-1}}, respectively. Circular markers indicate the 29 engineered TRGC variants strategically positioned to sample different regions of the theoretical parameter space. (f) Frequency response curves for all strains shown in (e) at $D = 0.3$. Solid lines represent theoretical predictions; data points show experimental measurements, demonstrating consistent agreement across diverse parameter combinations. (g) Frequency response characterization of selected TRGC variants at $D = 0.1$, showing distinct High-Pass FAC behaviors. Results validate maintenance of frequency discrimination capabilities at low duty cycles while demonstrating systematic variation in response characteristics based on molecular parameters. (h) Detailed characterization of strain FAC03C17V17 across multiple duty cycles, revealing systematic modulation of frequency response profiles. Lower duty cycles enhance frequency discrimination with reduced output amplitude, while higher duty cycles produce stronger outputs with modified frequency sensitivity. (i) Dynamic switching between high-pass and low-pass FAC behaviors in strain FAC03C22V34 through coordinated manipulation of light intensity and duty cycle. (j) Phase diagram mapping of transitions observed in (i), with solid circular points and arrows indicating parameter-driven switches between high-pass and low-pass regimes. Responses are evaluated at \SI{1.00e-2}{s^{-1}} and \SI{4.17e-4}{s^{-1}}. (k) Comprehensive correlation analysis between theoretical predictions and experimental measurements across the entire strain library under diverse conditions, validating the broad applicability of the theoretical framework ($R^2=0.986$). Error bars represent standard deviation from multiple experimental batches.}
\end{figure}

\newpage
\begin{figure}
	\centering
	\includegraphics[width=1.0\textwidth]{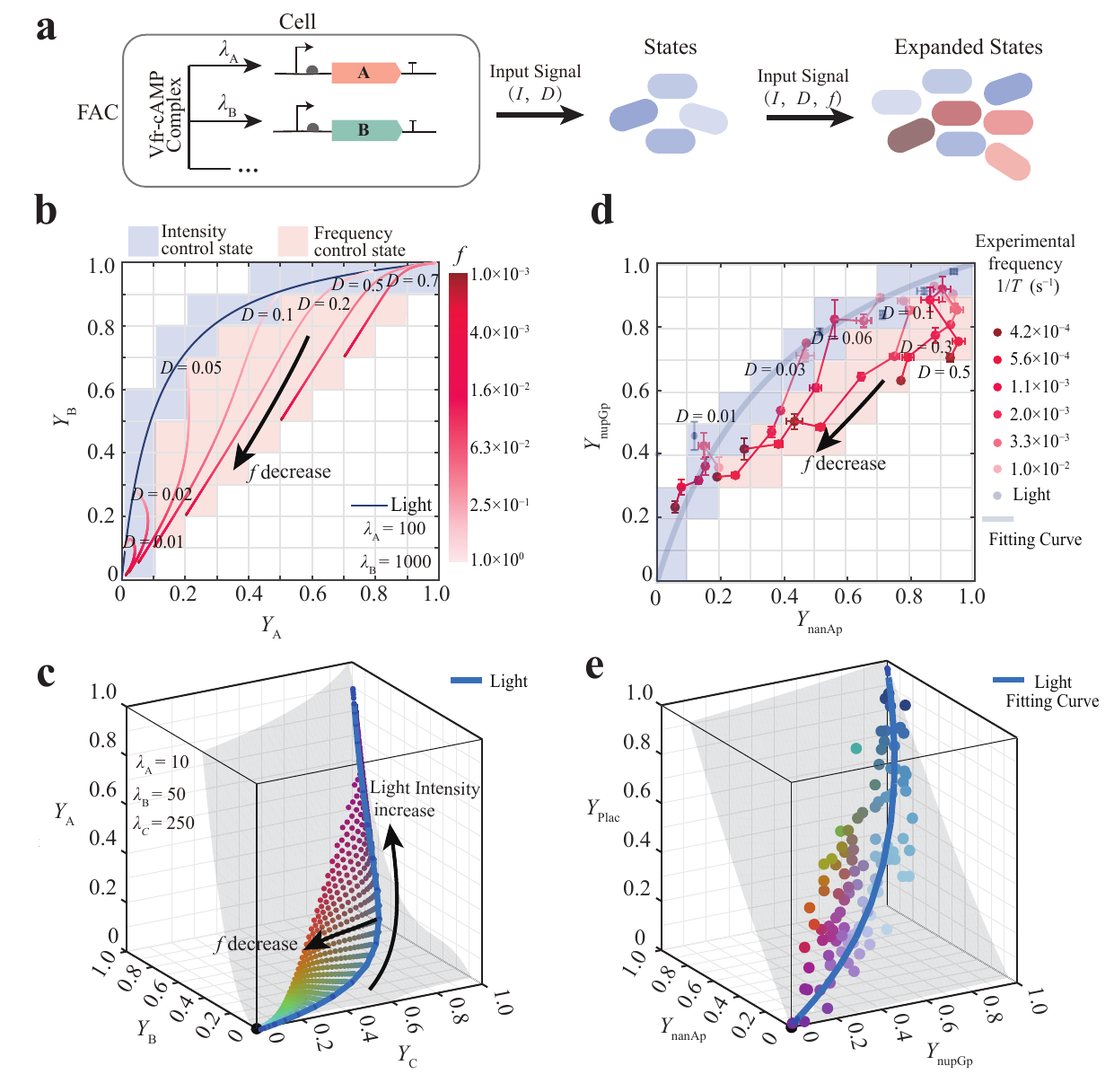}
	
	\raggedright
	\caption{\textbf{Frequency modulation expands accessible gene expression state combinations.}  \
		\normalfont (a) Schematic representation showing state space expansion achieved through frequency control as an additional parameter beyond amplitude modulation. (b) Theoretical analysis of a two-promoter system ($\epsilon = 0.1$ discretization). Blue curve and corresponding grid cells represent states accessible through amplitude modulation alone (19 states). Red curves and additional grid cells demonstrate state space expansion through frequency modulation (frequency range: $1$ to $1\times 10^{-3}$, non-dimensionalized units), achieving 38 total distinct states. (c) Three-dimensional state space simulation showing expansion from 27 states under amplitude modulation to 95 distinct states with frequency control. Color gradient represents transition from amplitude-only (blue) to frequency-modulated (red) states. Error bars represent standard deviation from three independent experiments. (d) Experimental validation using a two-promoter system. Normalized expression outputs ($Y$ values), derived from sfGFP and CyOFP fluorescence measurements, demonstrate predicted state space expansion through frequency modulation. Color scheme corresponds to theoretical predictions in (b). (e) Three-promoter experimental system showing normalized expression outputs ($Y$ values) obtained from sfGFP, CyOFP, and mScarlet fluorescence measurements, validating theoretical predictions from (c). Color scheme matches theoretical predictions in (c).}
\label{fig4}
\end{figure}

\end{document}